\documentclass[a4paper, 11pt]{article}
\usepackage[cmex10]{amsmath}
\usepackage{amssymb}
\usepackage{mathrsfs}
\usepackage{amsthm}
\usepackage{soul}
\usepackage{graphicx}
\usepackage{url}
\usepackage{cite}
\usepackage{fancyhdr}
\usepackage{makeidx}
\usepackage{authblk}
\usepackage{sectsty}
\sectionfont{\centering}
\subsectionfont{\centering}
\newtheorem{Theorem}{Theorem}
\newtheorem{Example}{Example}[section]

\newtheorem{Definition}{Definition}[section]
\newtheorem{Lemma}{Lemma}

\newtheorem{Corollary}{Corollary}

\newcommand{\pr} {{\bf Proof. \hspace{0.5cm}}}

\date{}

\begin{document}

\centerline{}

\centerline{}

\centerline {\Large{\bf Construction of DNA codes using $\theta$-skew cyclic codes over $\mathbb{F}_4 + v \mathbb{F}_4$}}

\centerline{}

\centerline{\bf {Joël Kabore, Mohammed Elhassani Charkani}}

\centerline{}

\centerline{}

\centerline{\bf Abstract}
{In this paper, we investigate $\theta$-skew cyclic codes over the ring $R= \mathbb{F}_4 + v \mathbb{F}_4$, where $v^2=v$ and $\theta$ is a non-trivial automorphism over $\mathbb{F}_4 + v \mathbb{F}_4$. This allows us to describe DNA code over this ring by characterizing $\theta$-skew cyclic reversible DNA codes and $\theta$-skew cyclic reversible complement DNA codes. We also explore the Gray images of $\theta$-skew cyclic codes.}

{\bf Keywords:}  \emph{$\theta$-Skew cyclic codes, reversible codes, DNA codes.}

\vspace{0.15cm}

\emph{AMS Subject Classification}: 94B05, 94B60, 11T71.

\section{Introduction}
Algebraic coding theory is an important branch of mathematics which plays an important role in information theory. Recently some connections between coding theory and biology, in particular in genetic with DNA molecules are established and the study of DNA codes have received a great attention. DNA computing has become an important research in coding theory \cite{AGZ, A, DVI, GG}.

DNA is formed by strands and each strand is a finite sequence consists of four nucleotides which are: Adenine(A), Guanine(G), Thymine(T), and Cytosine(C). The ends of a DNA strand are chemically polar with the so-called $5^{\prime}$ end and the $3^{\prime}$ ends. Two strands are linked with Watson-Crick Complement(WCC) which is characterized by $\overline{A}= T;~ \overline{T}=A;~ \overline{G}= C;~\overline{C}=G$ and Switching the $3^{\prime}$ and the $5^{\prime}$ end. In genetic, a DNA molecular is presented as a string of two complementary strands wound around each other. This process where a strand and its WCC bond to form a double helix is called hybridization.

A DNA code of length $n$ is simply defined as a nonempty subset of $\{A,~C,~G,~T\}^{n}$ which must satisfy some constraints. The most constraints used in DNA codes are: the Hamming distance constraint, the reverse constraint, the reverse complement constraint and the fixed GC-content constraint.

Adelman was the pioneer in studying DNA computing in 1994 \cite{A}. He used the WCC property of DNA strands to solve a  hard computational (NP) problem. DNA codes have been used in \cite{BDL}  to break the data encryption DES cryptographic system. 

Skew cyclic codes have been extensively developed  over finite chain rings \cite{BGU, JLU, SAS} and over some non-chain rings \cite{AAS, BR, GSY, KFGC}. 
In \cite{GOS}, Gursoy et al. have been used skew cyclic codes over finite fields to determine the structure of DNA codes. In \cite{BOS}, Bayram et al. explored skew cyclic codes over the ring $R= \mathbb{F}_4 + v \mathbb{F}_4$, using the non-trivial automorphism 
$\psi: R \rightarrow R$ such that $\psi(a + bv)= a^{2} + vb^{2}$; and defined a set to describe DNA codes. In our work, we use the non-trivial automorphism  $\theta: R \rightarrow R$ such that $\theta(a + bv)= a +b(1+v)$ to describe the structure of skew cyclic codes and to characterize DNA codes. This automorphism is more suitable to characterize DNA codes using skew cyclic codes. 

The remainder of this paper is organized as follows. In Section 2, we give some elementary definitions and properties about codes and DNA codes. In Section 3, we describe the structure of $\theta$-skew cyclic codes over $R$. In Section 4, we characterize DNA codes over $R$. Some necessary and sufficient conditions for the existence of reversible and reversible complement DNA codes over $R$ are given. In Section 5, we determine the Gray images of $\theta$-skew cyclic codes over $R$

\section{Preliminaries}
Let $\mathbb{F}_4:=\{0,1,\alpha,\alpha+1\}$ be the finite field with four elements, where $\alpha = \alpha^2+1$ and $R$ be the commutative ring $\mathbb{F}_4 + v \mathbb{F}_4$ with $v^2= v$.\\
The ring $R$ is a finite non chain ring and have two maximals ideals: $\langle v \rangle$ and $\langle v+1 \rangle$. By Chinese Remainder Theorem, we have: 
$$R= \langle v \rangle + \langle v+1 \rangle \cong R/ \langle v \rangle \times  R/ \langle v+1 \rangle $$
Any element $r$ in $R$ can be uniquely expressed as
$r= a+ bv=(a+b)v + a(v+1)$ where $a,b \in \mathbb{F}_4$. It is a unit if and only if $a\neq 0$ and $a+b \neq 0$. 

\begin{Lemma}\label{unit}
Let $\lambda= a +bv$ be a unit in $R$. Then $\lambda^{-1}= a^{-1} + b^{2}v$.
\end{Lemma}

The Gray map over R is defined by:
$$
\begin{array}{c c c c}
\phi:&  R & \rightarrow &\mathbb{F}_4^{2}\\
     &  a+ bv & \mapsto & (a + b, a)
\end{array}
$$
This map is linear and naturally extended to $R^n$ component-wise.
A code $C$ of length $n$ over $R$ is a nonempty subset of
$R^n.$ If in addition the code is a $R$-submodule of $R^n$, it
is called linear code.
The Hamming weight of a codeword $c=(c_0, \cdots, c_{n-1})$ denoted by $W_H(c)$ is the number of non-zeros components $c_i$, for all $0 \leq i \leq n-1$. The Hamming distance between two codewords $a, b$ is defined as 
$d_H(a,b)=W_H(a-b)$.

The four DNA nucleotides $A, T, C, G$ are respectively associated to the four elements $0, 1, \alpha, \alpha^2$ of $\mathbb{F}_4$. By the Watson-Crick Complement(WCC), 
we have: $\overline{0}= 1;~ \overline{1}= 0;~ \overline{\alpha}= \alpha^2;~ \overline{\alpha^2}= \alpha.$  

By using the Gray map $\phi$, a one-to-one correspondence $\Phi$ between elements of $R$ and DNA $2$-bases\\
$D = \{AA, AT, AG, AC, TT, TA, TG, TC, GG, GA, GC, GT, CC, CA, CG, CT \}$ is provided in Table \ref{cor}. Note that any $a \in R$ satisfies $a + \bar{a}= 1$.

\begin{table}[!h]
\begin{center}
\caption{correspondence $\Phi$ between elements of R and DNA 2-bases}
\label{cor}
\begin{tabular}{c c c } 
 \hline
 Elements of $R$ & Gray images &  DNA 2- bases\\  
 \hline
 $0$ & $(0,0)$ & AA \\ 
 $1$ & $(1,1)$ & TT \\
$\alpha$ & $(\alpha, \alpha)$ & CC \\
 $\alpha^2$ & $(\alpha^2, \alpha^2)$ & GG\\
 $v$ & $(1, 0)$ & TA  \\
$1+v$ & $(0, 1)$ &AT  \\ 
 $\alpha +v$ & $(\alpha^2, \alpha)$ & GC \\
 $\alpha^2 + v$ & $(\alpha, \alpha^2)$ & CG \\
 $\alpha v$ & $(\alpha, 0)$& CA\\
 $1+ \alpha v$ & $(\alpha^2, 1)$ & GT\\
$\alpha + \alpha v$ & $(0, \alpha)$& AC \\ 
 $\alpha^2+ \alpha v$ & $(1, \alpha^2)$ & TG\\
 $\alpha^2v$ & $(\alpha^2, 0)$ & GA \\
  $1 + \alpha^2v$  & $(\alpha, 1)$ & CT \\
 $\alpha + \alpha^2v$ & $(1, \alpha)$ & TC\\ 
$\alpha^2 + \alpha^2v$ & $(0, \alpha^2)$ & AG\\
\hline
\end{tabular}
\end{center}
\end{table}

For any codeword $u= (u_0,u_1,...,u_{n-1})$, we define the reverse of $u$ by $u^r= (u_{n-1},u_{n-2},...,u_{0})$; the complement of $u$ by $u^{c}=  (\overline{u}_0,\overline{u}_1,...,\overline{u}_{n-1})$ and the reverse complement by $u^{rc}= (\overline{u}_{n-1},\overline{u}_{n-2},...,\overline{u}_{0})$.

\begin{Definition}
A linear code $\mathcal{C}$ of length $n$ over $R$ is said to be:
\begin{enumerate}
\item reversible if the reverse of each codeword is also in $\mathcal{C}$;
\item complement if the complement of each codeword is also in $\mathcal{C}$;
\item reversible complement if the reverse complement of each codeword is also in $\mathcal{C}$.
\end{enumerate}
\end{Definition}

\section{Structure of $\theta$-skew cyclic codes over $R$}

Let $\Theta$ be an automorphism  over $R$. The skew polynomial ring denoted by $R[x, \Theta]$ is defined as the set
$R[x]$ of formal polynomials over $R$, where the addition is
the usual addition of polynomials and the
multiplication is defined  using the rule  $x a = \Theta(a)x$ (
$a$ in $R$), which is extended to all elements of $R[x,
\Theta]$ by associativity and distributivity.\\
A $\Theta$-skew cyclic shift $\sigma_{\Theta}$ is defined  on $R^n$ by:
$$\sigma_{\Theta}(a_0,a_1,\cdots,a_{n-1})=(\Theta(a_{n-1}), \Theta(a_0),\cdots, \Theta(a_{n-2})).$$
A linear code $C$ of length $n$ which is invariant under a  $\Theta$-skew cyclic
shift $\sigma_{\Theta}$ is called $\Theta$-skew cyclic code. It is well known that
$\Theta$-skew cyclic codes of length $n$ over
$R$ can be identified with the left $R[x,\Theta]$-submodule of
$R_n:= R[x, \Theta]/\langle x^n-1 \rangle$ by the
identification:
$$ \varphi: (a_0,a_1,\cdots,a_{n-1})\longmapsto a_0+a_1x+\cdots+a_{n-1}x^{n-1}$$
and  the multiplication defined by
$$f(x)\left(g(x)+ \langle\,x^n- 1\,\rangle\right)= f(x)g(x)+ \langle\,x^n- 1\,\rangle,$$
with $f(x), g(x) \in R[x, \Theta]$.\\

Let $\theta$ be the automorphism over $R$ defined by:
\begin{align}
\begin{array}{cccc}
\theta: & R& \rightarrow & R\\
&a + bv &\mapsto & a + b(1 +v).
\end{array}
\end{align}
Note that $\theta$ is a ring automorphism of order $2$.

\begin{Lemma}
Let $\mathcal{C}$ be a $\theta$-skew cyclic code over $R$ and $g$ be a polynomial in $\mathcal{C}$ of minimal degree such that the leading coefficient of $g$ is not a unit then $g(x)= v g_1(x)$ or $g(x)=(v+1) g_1(x)$ where $g_1 \in \mathbb{F}_4[x]$.
\end{Lemma}
\pr
The proof is similar to Lemma 2 in \cite{AAS}.
$\qed$

\begin{Theorem}
Let $\mathcal{C}$ be a $\theta$-skew cyclic code of length $n$ over $R$. Then
\begin{enumerate}
\item If there exists a polynomial $g(x)$ in $\mathcal{C}$ of minimal degree such that its leading coefficient is a unit in $R$, then $\mathcal{C}= \langle g(x) \rangle,$ where $g(x)$ is a right divisor of $x^n-1$.
\item Suppose there is no polynomial in $\mathcal{C}$ of minimal degree with leading coefficient a unit. Then
\begin{enumerate}
\item If there exists some polynomials in $\mathcal{C}$ with leading coefficient a unit, then $\mathcal{C}= \langle f(x), g(x) \rangle,$ where $g(x)$ is a polynomial of minimal degree amongst polynomials in $\mathcal{C}$ with leading coefficient a unit and $f(x)$ is a polynomial in $\mathcal{C}$  of minimal degree with leading coefficient a non-unit.
\item If there is no polynomial in $\mathcal{C}$  with leading coefficient a unit, then $\mathcal{C}= \langle f(x) \rangle,$ where $f(x)$ is a polynomial in $\mathcal{C}$  of minimal degree with leading coefficient a non-unit.\\
Moreover $f(x)= vf_1(x)$ or $f(x)= (v+1)f_1(x)$, where $f_1(x)$ divides $x^n-1$ in $\mathbb{F}_4[x]$.
\end{enumerate}
\end{enumerate}
\end{Theorem}

\pr
The proof is similar to Corollary 5 in \cite{AAS}.
$\qed$

\section{Reversible and reversible-complement DNA codes over $R$}

In this section we characterize the generators of $\theta$-skew cyclic reversible and $\theta$-skew cyclic reversible-complement codes over $R$.
The following definition can be found in \cite{GOS}.

\begin{Definition}
Let $f(x)= a_0 + a_1 x+ \cdots + a_t x^t$ be a polynomial of degree $t$ over $R$ and $\theta$ be an automorphism over $R$. The polynomial $f(x)$ is said to be a palindromic polynomial if $a_i=a_{t-i}$ for all $i \in \{0, 1 \cdots, t\}$ and $f(x)$ is said to  be a $\theta$-palindromic polynomial if $a_i=\theta(a_{t-i})$ for all $i \in \{0, 1 \cdots, t\}$.
\end{Definition}

Let $\mathcal{C}$ be a code over $R$ and $c$ be a codeword.
From the correspondence $\Phi$, it is easy to see that the reverse of $\Phi(c)$ is given by $\Phi(c)^r= \Phi(\theta(c)^r).$

\begin{Definition}
Let $\mathcal{C}$ be a code of length $n$ over $R$. If $\Phi(c)^r \in \Phi(\mathcal{C})$ for all $c \in \mathcal{C}$, then $\mathcal{C}$ or equivalently $\Phi(\mathcal{C})$ is called a reversible DNA code. Similarly $\mathcal{C}$ or equivalently $\Phi(\mathcal{C})$ is called a reversible-complement DNA code if $\Phi(c)^{rc} \in \Phi(\mathcal{C})$ for all $c \in \mathcal{C}$.
\end{Definition}

\begin{Theorem}
Let $\mathcal{C}= \langle g(x) \rangle$ be a $\theta$-skew cyclic code of even length $n$ over $R$, where $\deg (g(x))$ is even.

Then $\mathcal{C}$ is a reversible DNA code if and only if $g(x)$ is a palindromic polynomial.
\end{Theorem}
\pr
Let $g(x)= a_0 + a_{1}x+ \cdots +a_1x^{t-1}+ a_0 x^{t}$ be a palindromic polynomial of degree $t$ and $c$ be an element of $\mathcal{C}$. Since the reverse of the DNA codeword $\Phi(c)$ is given by $\Phi((\theta(c))^r);$ then:
$$(\Phi(\sum\limits_{i=0}^{n-t-1}\lambda_i x^i g(x)))^r= \Phi(\sum\limits_{i=0}^{n-t-1} \theta(\lambda_i) x^{n-t-1-i} g(x));$$
where $\lambda_i \in R$.
Since $\mathcal{C}$ is a  $\theta$-skew cyclic code, then $\mathcal{C}$ is a reversible DNA code.

Conversely, suppose $\mathcal{C}= \langle g(x) \rangle$ is a reversible $\theta$-skew cyclic code.
\begin{itemize}
\item [$i)$] If the leading coefficient of $g$ is a unit, then, without loss of generality, we can suppose $g$ is monic. Since $g(x)= a_{0} + a_{1}x+ \cdots  +a_{t-1}x^{t-1}+ x^{t}$ is a monic right divisor of $x^n-1$ in $R[x,\theta]$, then $a_0$ is a unit in $R$. Since $\mathcal{C}$ is a reversible DNA code, then $g^r(x)= x^{n-t-1}+ \theta(a_{t-1}) x^{n-t}+ \cdots + \theta(a_{1})x^{n-2} + \theta(a_{0})x^{n-1} \in \mathcal{C}$. Moreover, 
$x^{t+1}g^r(x)=1 + a_{t-1}x+ \cdots+ a_1 x^{t-1}+ a_0 x^{t} \in \mathcal{C}$. We have 
$g(x)- a_0^{-1} x^{t+1}g^r(x)= (a_0 - a_0^{-1})+ (a_{1} - a_0^{-1} a_{t-1})x + \cdots+ (a_{t-1} - a_0^{-1} a_{1})x^{t-1} \in \mathcal{C}$. By minimality of $\deg(g)$, we have $g(x)- a_0^{-1} x^{t+1}g^r(x)=0$. This implies that \\
$(a_0 - a_0^{-1})= (a_{1} - a_0^{-1} a_{t-1})= \cdots= 0$. Whence $ a_0= 1$ and $a_i=a_{t-i}$ for all $i \in \{0,1,\cdots,t\}$. It follows that $g(x)$ is a palindromic polynomial. 

\item[$ii)$] If $g(x)= v g_1(x)$ where $g_1(x)= a_{0} + a_{1}x+ \cdots + a_{t-1}x^{t-1}+ x^{t}$ is a monic divisor of $x^n-1$ in $\mathbb{F}_4[x]$, then $a_0$ is a unit in $\mathbb{F}_4$.
Since $\mathcal{C}$ is a reversible DNA code, then $g^r(x)= (v+1)(x^{n-t-1}+ a_{t-1} x^{n-t}+ \cdots + a_{1}x^{n-2} + a_{0}x^{n-1}) \in \mathcal{C}$. Moreover
$x^{t+1}g^r(x)= v(1 + a_{t-1}x+ \cdots+ a_1 x^{t-1}+ a_0 x^{t}) \in \mathcal{C}$. We have 
$g(x) - a_0^{-1} x^{t+1}g^r(x)=v (a_0- a_0^{-1}) + v (a_1- a_0^{-1} a_{t-1})x + \cdots v(a_{t-1} - a_0^{-1}a_1)x^{t-1} \in \mathcal{C}$.

By minimality of $\deg(g)$, we have $g(x) - a_0^{-1} x^{t+1}g^r(x)=0$, that's to say\\
$(a_0- a_0^{-1})= (a_1- a_0^{-1} a_{t-1})= (a_{t-1} - a_0^{-1}a_1)= 0$. Whence $ a_0= 1$ and $a_i=a_{t-i}$ for all $i \in \{0,1,\cdots,t\}$.

The proof is similar if $g(x)= (v+1) g_1(x)$ where $g_1(x)= a_{0} + a_{1}x+ \cdots + a_{t-1}x^{t-1}+ x^{t}$ is a monic divisor of $x^n-1$ in $\mathbb{F}_4[x]$.
\end{itemize}
$\qed$

\begin{Example}
The polynomial $g(x)= x^4+ (v+ \alpha)x^2+1$ is a right divisor of $x^{10} -1$ over $R[x,\theta]$ and is a palindromic polynomial. Then the $\theta$-skew cyclic code generated by $g(x)$ is a reversible DNA code of length $10$ over $R$.
\end{Example}

\begin{Theorem}\label{palin}

Let $\mathcal{C}= \langle g(x) \rangle$ be a $\theta$-skew cyclic code of even length $n$ over $R$, where $g(x)$ is a right divisor of $x^n-1$ in $R[x,\theta]$ with leading coefficient a unit. Suppose that $\deg (g(x))$ is odd.

Then $\mathcal{C}$ is a reversible DNA code if and only if  $\mathcal{C}$ is generated by a $\theta$-palindromic polynomial.
\end{Theorem}
\pr
Without loss of generality, we can suppose $g$ is monic.
Let $g(x)= 1 + a_{1}x+ \cdots +\theta(a_1)x^{t-1}+ x^{t}$ be a $\theta$-palindromic polynomial and $c$ be an element of $\mathcal{C}$. Since the reverse of the DNA codeword $\Phi(c)$ is given by $\Phi((\theta(c))^r);$ then:
$$(\Phi(\sum\limits_{i=0}^{n-t-1}\lambda_i x^i g(x)))^r= \Phi(\sum\limits_{i=0}^{n-t-1} \theta(\lambda_i) x^{n-t-1-i} g(x));$$
where $\lambda_i \in R$.
Since $\mathcal{C}$ is a  $\theta$-skew cyclic code, then $\mathcal{C}$ is a reversible DNA code.

Conversely, suppose $\mathcal{C}= \langle g(x) \rangle$ is a reversible $\theta$-skew cyclic code. Let  
$g(x)= a_{0} + a_{1}x+ \cdots  +a_{t-1}x^{t-1}+ x^{t}$. Note that $g$ divides $x^n-1$ implies that $a_0$ is a unit in $R$. Since $\mathcal{C}$ is a reversible DNA code, then $g^r(x)= x^{n-t-1}+ \theta(a_{t-1}) x^{n-t}+ \cdots + \theta(a_{1})x^{n-2} + \theta(a_{0})x^{n-1} \in \mathcal{C}$. Moreover, 
$x^{t+1}g^r(x)=1 + \theta(a_{t-1})x+ \cdots+ \theta(a_1) x^{t-1}+ \theta(a_0) x^{t} \in \mathcal{C}$. We have 
$g(x)- \theta(a_0^{-1}) x^{t+1}g^r(x)= (a_0 - \theta(a_0^{-1}))+ (a_{1} - \theta(a_0^{-1})\theta(a_{t-1}))x + \cdots+ (a_{t-1} - \theta(a_0^{-1}) \theta(a_{1}))x^{t-1} \in \mathcal{C}$. By minimality of $\deg(g)$, we have $g(x)- \theta(a_0^{-1}) x^{t+1}g^r(x)=0$. This implies that  $(a_0 - \theta(a_0^{-1}))= (a_{1} - \theta(a_0^{-1}) \theta(a_{t-1}))= \cdots= 0$. \\
From Lemma \ref{unit}, we deduce that
$ a_0 = \theta(a_0^{-1})$ implies that $a_0 \in \{1, \alpha +v, \alpha^2 + v\}$. It is easy to see that $a_0^{2}= \theta(a_0)$.\\
Then $g(x)= \sum\limits_{i=0}^{\frac{t-1}{2}}( a_i x^{i} + a_0 \theta(a_{i}) x^{t-i})$ and 
$a_0 g(x)=  \sum\limits_{i=0}^{\frac{t-1}{2}}( a_0 a_i x^{i} + \theta(a_0 a_{i}) x^{t-i}) \in \mathcal{C}$.
The result follows by the fact that $\mathcal{C}= \langle g(x) \rangle= \langle a_0 g(x) \rangle.$

$\qed$

\begin{Example}
The polynomial $g(x)= x^3+ (v+ \alpha^2)x^2+(v+ \alpha)x+1$ is a right divisor of $x^{12} -1$ over $R[x,\theta]$ and is a $\theta$-palindromic polynomial. Then the skew cyclic code generated by $g(x)$ is a reversible DNA code of length $12$ over $R$
\end{Example}

Suppose $\mathcal{C}$ is a $\theta$-skew cyclic code of odd length $n$ over $R$ and $c=(c_0,c_1, \cdots, c_{n-1})$ be a codeword. Then $(\theta(c_{n-1}),\theta(c_0), \cdots, \theta(c_{n-2})) \in \mathcal{C}$. Since $n$ is odd, we have $(c_{n-1},c_0, \cdots, c_{n-2}) \in \mathcal{C}$. That's to say $\mathcal{C}$ is cyclic.

\begin{Theorem}
Let $\mathcal{C}= \langle g(x) \rangle$ be a $\theta$-skew cyclic code of odd length $n$ over $R$, where $g(x)$ is a right divisor of $x^n-1$ in $R[x,\theta]$ with leading coefficient a unit. Then 
\begin{enumerate}
\item If $g(x)$ is palindromic or $\theta$-palindromic, then $\mathcal{C}$ is a reversible DNA code.
\item If $\mathcal{C}$ is a reversible DNA code, then $\mathcal{C}$ is a cyclic code over $R$ generated by a palindromic polynomial $\tilde{g}(x) \in \mathbb{F}_4[x]$. 
\end{enumerate}
\end{Theorem}
\pr
Without loss of generality, we can suppose that $g$ is monic.
We suppose that $t$ is odd.\\
Let $g(x)= a_{0} + a_{1}x+ \cdots  +a_{t-1}x^{t-1}+ x^{t}$. Since $n$ is odd, then 
$g_{\theta}(x)= x^{n}g(x)= \theta({a}_{0}) + \theta({a}_{1})x+ \cdots  + \theta({a}_{t-1})x^{t-1}+ x^{t} \in \mathcal{C}$.
If $g$ is palindromic
then:
$$(\Phi(\sum\limits_{i=0}^{n-t-1}\lambda_i x^i g(x)))^r= \Phi(\sum\limits_{i=0}^{n-t-1} \theta(\lambda_i) x^{n-t-1-i} g(x));$$
where $\lambda_i \in R$.

If $g$ is $\theta$-palindromic,
then:
$$(\Phi(\sum\limits_{i=0}^{n-t-1}\lambda_i x^i g(x)))^r= \Phi(\sum\limits_{i=0}^{n-t-1} \theta(\lambda_i) x^{n-t-1-i} g_{\theta}(x));$$
where $\lambda_i \in R$.
Since $\mathcal{C}$ is $\theta$-skew cyclic, the result follows.

Conversely, suppose $\mathcal{C}= \langle g(x) \rangle,$ is a reversible $\theta$-skew cyclic code, where $g(x)= a_{0} + a_{1}x+ \cdots  +a_{t-1}x^{t-1}+ x^{t}$ is a monic right divisor of $x^n-1$ in $R[x,\theta]$ with odd degree.

We have $g^r(x)= x^{n-t-1}+ \theta(a_{t-1}) x^{n-t}+ \cdots + \theta(a_{1})x^{n-2} + \theta(a_{0})x^{n-1} \in \mathcal{C}$ and $x^{t+1}g^r(x)=1 + \theta(a_{t-1})x+ \cdots+ \theta(a_1) x^{t-1}+ \theta(a_0) x^{t} \in \mathcal{C}$. By a similar work as in the proof of Theorem \ref{palin}; we deduce that $\mathcal{C}$ is generated by a $\theta$-palindromic polynomial $\tilde{g}(x)= b_{0} + b_{1}x+ \cdots  +\theta(b_1) x^{t-1}+ \theta(b_0)x^{t}$.

Since $n$ is odd, then $\tilde{g}_{\theta}(x)= \theta(b_0) + \theta(b_1)x+ \cdots  + b_1 x^{t-1}+ b_0 x^{t} \in \mathcal{C}$. This implies that 
$\theta(b_0^{-1}) \tilde{g}(x)- b_0^{-1}\tilde{g}_{\theta}(x)= (\theta(b_0^{-1}) b_{0}-b_0^{-1}\theta(b_0)) +(\theta(b_0^{-1}) b_{1}-b_0^{-1}\theta(b_1))x+ \cdots  + (\theta(b_0^{-1}) \theta(b_{1})-b_0^{-1}b_1) x^{t-1}=0$
by minimality of the degree of $\tilde{g}$. Whence 
$
(\theta(b_0^{-1}) b_{0}-b_0^{-1}\theta(b_0))= (\theta(b_0^{-1}) b_{1}-b_0^{-1}\theta(b_1))= \cdots = (\theta(b_0^{-1}) \theta(b_{1})-b_0^{-1}b_1)=0.$

We have $\theta(b_0^{-1}) b_{0}-b_0^{-1}\theta(b_0)=0$ implies that $\theta(b_0^{-1}) b_{0}=1$, that is to say $b_{0}= \theta(b_0)$. Then $b_{0} \in \mathbb{F}_4$. It follows that $b_{i}=\theta(b_i)$ that is say $b_i \in \mathbb{F}_4$ for all  $i \in \{0,1,\cdots t\}$.

Similar arguments work if $t$ is even.

$\qed$

\begin{Theorem}
Let $\mathcal{C}= \langle v g_1(x) \rangle$ or  $\mathcal{C}= \langle (v+1)g_1(x) \rangle$ be a $\theta$-skew cyclic code of length $n$ over $R$, where $g_1(x)$ is a divisor of $x^n-1$ in $\mathbb{F}_4[x]$. Suppose  $n$ is odd or ($n$ is even and $\deg(g(x))$ is odd).
Then $\mathcal{C}$ cannot be a reversible DNA code.
\end{Theorem}
\pr

Let  $\mathcal{C}= \langle g(x) \rangle,$ where $g(x)= vg_1(x)$ and $g_1(x)= a_{0} + a_{1}x+ \cdots + a_{t-1}x^{t-1}+ x^{t}$ be a divisor of $x^n-1$ in $\mathbb{F}_4[x]$.

If $n$ is odd, then $x^{n}v g_1(x)=(v+1)g_1(x) \in \mathcal{C}$, this implies that $g_1(x) \in \mathcal{C}$, which is not possible.

Suppose $n$ is even  and $\deg(g)$ is odd.
If $\mathcal{C}$ is a reversible DNA code, then $g^r(x)=(v+1)(x^{n-t-1}+ a_{t-1} x^{n-t}+ \cdots + a_{1}x^{n-2} + a_{0}x^{n-1}) \in \mathcal{C}$. 
Moreover, 
$x^{t+1}g^r(x)= (v+1)(1 + a_{t-1} x+ \cdots+ a_1 x^{t-1}+ a_0 x^{t}) \in \mathcal{C}$ 
implies that 
$g(x)+ a_0^{-1} x^{t+1}g^r(x)= x^t+ (v a_{t-1}+ (v+1) a_0^{-1}a_1)x^{t-1}+ \cdots+ v a_0 + (v+1) a_0^{-1}a_t \in \mathcal{C}$.
This contradicts the fact that $\mathcal{C}$ has no polynomials in $\mathcal{C}$ with leading coefficient a unit. Then $\mathcal{C}$ cannot be a reversible code.

$\qed$

Let $\mathcal{C}$ be a $\theta$-skew cyclic code of length $n$ over $R$ and $c(x)= a_{0} + a_{1}x+ \cdots+ a_{n-1}x^{n-1}$ be a codeword. If $\mathcal{C}$ is complement then $c^{c} \in \mathcal{C}$ and $c^{c}(x)+ c(x)= 1 + x+ \cdots+ x^{n-1} \in \mathcal{C}$. Then, we characterize reversible complement DNA codes over $R$ as follows.

\begin{Corollary}
Let $\mathcal{C}= \langle g(x) \rangle$ be a $\theta$-skew cyclic code of length $n$ over $R$, where $g(x)$ is a right divisor of $x^n-1$ in $R[x,\theta]$.
\begin{enumerate}
\item If $n$ and $\deg(g(x))$ are even, then $\mathcal{C}$ is reversible complement DNA code if and only if $g(x)$ is a palindromic polynomial and $1 + x+ \cdots+ x^{n-1} \in \mathcal{C}$.  
\item If $n$ is even and $\deg (g(x))$ is odd, then $\mathcal{C}$ is a reversible complement DNA code if and only if $\mathcal{C}$  is generated by a $\theta$-palindromic polynomial and $1 + x+ \cdots+ x^{n-1} \in \mathcal{C}$.
\item If $n$ is odd, then $\mathcal{C}$ is a reversible complement DNA code if and only if $\mathcal{C}$ is a cylic code over $R$ generated by a polynomial $\tilde{g}(x) \in \mathbb{F}_4[x]$ and $1 + x+ \cdots+ x^{n-1} \in \mathcal{C}$
\end{enumerate}
\end{Corollary}

\begin {Corollary}
Let $\mathcal{C}= \langle v g_1(x) \rangle$ or  $\mathcal{C}= \langle (v+1)g_1(x) \rangle$ be a $\theta$-skew cyclic code of length $n$ over $R$, where $g_1(x)$ is a divisor of $x^n-1$ in $\mathbb{F}_4[x]$.
Then the skew cyclic code  $\mathcal{C}$ cannot be a complement DNA code.
\end{Corollary}

\section{Gray images of $\theta$-skew cyclic codes over $R$}

The Lee weight of an element $a$ in $R$ is defined as $W_L(a)= W_H( \Phi(a))$ and the Lee distance between two elements in $R$ is given by $d_L(a,b)= W_L(a-b)$.
The following result is straightforward.

\begin{Lemma}
The Gray map: $\Phi: (R^{n}, d_L) \longrightarrow (F_4^{2n}, d_H)$ is a distance-preserving map.
\end{Lemma}

Let $\mathcal{C}$ be a code of length $2n$ over $\mathbb{F}_4$ and $c=(a_0,b_0, a_1, b_1, \cdots, a_{n-1},b_{n-1})$ be a codeword. The code $\mathcal{C}$ is called a $2$-quasi cyclic code if it is invariant under the $2$-quasi cyclic shift defined by $\tau_2(c)=(a_{n-1},b_{n-1}, a_1, b_1, \cdots, a_{n-2},b_{n-2})$.

\begin{Theorem}
Let $\mathcal{C}$ be a $\theta$-skew cyclic code of length $n$ over $R$. Then $\Phi(\mathcal{C})$ is permutation equivalent to a $2$-quasi cyclic code of length $2n$ over $\mathbb{F}_4$.
\end{Theorem}
\pr
Let $c=(a_0 + b_0 v, a_1 + b_1 v, \cdots, a_{n-1} + b_{n-1}v) \in \mathcal{C}$. We have  
\begin{eqnarray*}
\tau_2(\Phi(c)) &=& \tau_2((a_0 + b_0, a_0, a_1 + b_1, a_1, \cdots, a_{n-1} + b_{n-1}, a_{n-1})) \\
                &=& (a_{n-1} + b_{n-1}, a_{n-1}, a_0 + b_0, a_0, \cdots, a_{n-2} + b_{n-2}, a_{n-2}) 
\end{eqnarray*}

and

\begin{eqnarray*}
\Phi(\sigma_{\theta}(c)) &=& \Phi(\theta(a_{n-1} + b_{n-1}v), \theta(a_0 + b_0 v), \cdots, \theta(a_{n-2} + b_{n-2}v)) \\
                         &=& \Phi(a_{n-1} + b_{n-1} + b_{n-1}v, a_0 + b_0 + b_0 v, \cdots, a_{n-2} + b_{n-2}, b_{n-2}v) \\
												 &=&(a_{n-1}, a_{n-1} + b_{n-1}, a_0 ,a_0 + b_0, \cdots, a_{n-2}, a_{n-2} + b_{n-2})
\end{eqnarray*}

Whence, If $\mathcal{C}$ is a $\theta$-skew cyclic code of length $n$ over $R$, then $\Phi(\mathcal{C})$ is a $2$-quasi cyclic code of length $2n$ over $\mathbb{F}_4$.
$\qed$

\begin{Example}
Let $\mathcal{C}=<v(x^4+x^2+1)>$ be the skew cyclic of length $6$ over $R$. The polynomial $g(x)= v(x^4+x^2+1)$ is a palindromic polynomial and generates a reversible DNA code over $R$. The image of $\mathcal{C}$ under the Gray map $\Phi$ provides a reversible DNA code of length 12 and minimum distance 3. These codewords are given in Table \ref{dna}

\begin{table}[!h]
\begin{center}
\caption{Reversible DNA code of length $12$ obtained from $\mathcal{C}=<v(x^4+x^2+1)>$}
\label{dna}
\begin{tabular}{c c c c }  
 \hline
 \footnotesize$AAAAAAAAAAAA$ &\footnotesize $TAATTAATTAAT$ & \footnotesize$CAACCAACCAAC$ &\footnotesize $GAAGGAAGGAAG$  \\ 
 \footnotesize$TAAATAAATAAA$ & \footnotesize$AAATAAATAAAT$ &\footnotesize $CAAACAAACAAA$ &\footnotesize $AAACAAACAAAC$ \\
 \footnotesize$GAAAGAAAGAAA$& \footnotesize$AAAGAAAGAAAG$ & \footnotesize$CAATCAATCAAT$ & \footnotesize$TAACTAACTAAC$\\
 \footnotesize$GAATGAATGAAT$&\footnotesize $TAAGTAAGTAAG$ & \footnotesize$GAACGAACGAAC$ &\footnotesize $CAAGCAAGCAAG$\\
\hline
\end{tabular}
\end{center}
\end{table}
\end{Example}

\end{document}